\documentstyle[12pt,epsfig,amssymb]{article}
\begin{document}
\begin{flushright}
IHEP-99-55\\
\end{flushright}

\begin{center}

{\Large \bf The QCD renormalization scale stability 
of high twists and $\alpha_s$ in deep inelastic scattering}

\vspace{0.1in}

{\bf S.I. Alekhin}

\vspace{0.1in}
{\baselineskip=14pt Institute  
for High Energy Physics, 142281 Protvino, Russia}

\begin{abstract}
A sensitivity of twist-4  and $\alpha_{\rm s}$ 
values extracted in the NLO QCD analysis 
of nonsinglet SLAC-BCDMS-NMC deep inelastic scattering data
to the choice of QCD renormalization scale (RS) is analysed. 
It is obtained that the high twist (HT) contribution to 
structure function $F_2$, is retuned with the change of RS.
This retuning depends on the choice of starting QCD evolution 
point $Q_0$ and $x$. At $Q_0\gtrsim 10$~GeV$^2$ the HT contribution to $F_2$
is retuned at small $x$ and almost not retuned at 
large $x$; at small $Q_0$ it exhibits approximate RS stability for 
all $x$ in question. 
The HT contribution to $F_{\rm L}$ is RS stable for all $Q_0$ and $x$.
The RS sensitivity of $\alpha_{\rm s}$ also depends on the 
choice of $Q_0$: at large $Q_0$ this sensitivity is 
weaker, than at small one. For $Q_0^2=9$~GeV$^2$ the value 
$\alpha_{\rm s}\left(M_{\rm Z}\right)=0.1183\pm0.0021({\rm stat+syst})\pm0.0013({\rm RS})$
is obtained. Connection with the higher order QCD corrections is discussed.
\end{abstract}
\end{center}
\newpage

{\bf 1.} Interest to the quantitative description of high twist (HT)
contribution to the deep inelastic scattering (DIS) cross sections
has been increased recently, in particular,
due to the development of infrared renormalon (IRR) model 
(see e.g. review \cite{REW}). Within this model  
one can derive the $x$-shape of HT contribution 
from the $x$-shape of leading twist (LT) structure functions.
This connection  allows for to obtain precise predictions 
for the HT contribution 
since the LT contribution can be rather precisely determined 
from experimental data.
Meanwhile the experimental determination of HT contribution is not 
direct and is based on fitting a combination of log- and 
power-like terms to the data. If the data accuracy is not high 
enough, the correlation between these log- and power-like terms
can be large, that was explicitly shown in the 
combined SLAC-BCDMS data analysis of Ref.~\cite{ALS}.
 If the large correlations occur, the separation 
of terms is unstable with respect to the various inputs of 
fit and thus it becomes important to study the stability of  
HT separation. One of the poorly defined ansatz of a QCD 
analysis of DIS data is the choice of renormalization scale (RS). 
The uncertainty due to RS variation is connected with the 
account of higher order (HO) QCD corrections since in the analysis 
with complete account of HO terms,
the RS dependence of fitted parameters should vanish and 
thus the observed RS dependence can be used 
for the estimation of HO terms effect. Earlier we reported the 
results of NLO QCD analysis of high $x$ SLAC-BCDMS-NMC data \cite{FL}.
A short communication 
on the RS dependence of HT contribution and $\alpha_{\rm s}$ 
obtained in this analysis was reported in Ref.~\cite{MOR99}. 
In this paper more detailed study of this dependence is given.

{\bf 2.} Our approach used for the study of RS
stability of DGLAP evolution equation in NLO QCD is the same as 
described in Refs.~\cite{MRS,VM}. Within this approach the RS
of QCD evolution is changed 
from $Q$ to $k_{\rm R}Q$, where $k_{\rm R}$ is arbitrary parameter, conventionally 
varied from 1/2 to 2.
This approach contains certain simplification since the change 
of scale can depend on $x$, but  
in our analysis this effect is not so essential due to the limited 
range of $x$. For the nonsinglet case 
NLO DGLAP equation with an arbitrary choice of RS looks as follows:
\begin{equation}
Q\frac{\partial q^{\rm NS}(x,Q)}{\partial Q}=
\frac{\alpha_{\rm s}\left(k_{\rm R}Q\right)}{\pi}
P^{\rm NS,(0)}_{\rm qq}\otimes q^{\rm NS}
+\frac{\alpha_{\rm s}^2(k_{\rm R}Q)}{2\pi^2}\left [
P^{\rm NS,(1)}_{\rm qq}\otimes q^{\rm NS}
+\ln(k_{\rm R})\beta_0P^{\rm NS,(0)}_{\rm qq}\otimes q^{\rm NS}\right ]
\label{eqn:dglap}
\end{equation}
where $P\otimes q=\int_x^1dzP(z)q(x/z,Q)$ denotes Mellin convolution;
$q^{\rm NS}$ is the evolved distribution;
$P^{\rm NS,(0)}$ and $P^{\rm NS,(1)}$ are the LO and NLO parts of 
splitting function; $\alpha_{\rm s}(Q)$ is the running strong coupling constant;
and $\beta_0$ is the regular coefficient of renormgroup equation for 
$\alpha_{\rm s}$:
\begin{displaymath}
Q\frac{d\alpha_{\rm s}}{dQ}=-\frac{\beta_0}{2\pi}\alpha_{\rm s}^2
-\frac{\beta_1}{4\pi^2}\alpha_{\rm s}^3.
\end{displaymath}
The  equation (\ref{eqn:dglap}) was solved with the help of 
direct integration method implemented in the code used earlier \cite{ALS}. 
In the analysis of Ref.~\cite{FL}
we used combined SLAC-BCDMS-NMC proton-deuterium 
data \cite{DATA}
with the cuts $x\ge0.3$ to reduce QCD evolution to the nonsinglet case
and $x\le0.75$ to reject the region where a nuclear effects in deuterium 
can be significant. 
The initial scale of evolution was chosen equal to $Q^2_0=9~$GeV$^2$
to provide comparability with the earlier results of Ref.~\cite{ALS}. 
A complete account of point-to-point correlations 
due to systematic errors was made through the covariance matrix approach, 
similarly to our earlier analysis of Ref. \cite{AL96}.
The formula were fitted to the cross section 
data to allow for simultaneous and unbiased determination of 
a twist-4 contribution to structure functions $F_{\rm L}$ and $F_2$: 
\begin{displaymath}
\frac{d^2\sigma}{dxdy}=\frac{4\pi\alpha^2(s-M^2)}{Q^4}
\left[\left(1-y-\frac{(Mxy)^2}{Q^2}\right)F_2^{\rm HT}(x,Q)+\\
\left(1-2\frac{m_{\rm l}^2}{Q^2}\right)
\frac{y^2}{2}2xF_1^{\rm HT}(x,Q)\right],
\end{displaymath}
\begin{displaymath}  
2xF_1^{\rm HT}(x,Q)=F_2^{\rm HT}(x,Q)-F_{\rm L}^{\rm HT}(x,Q),
\end{displaymath}
\begin{displaymath}  
F_{2,\rm L}^{\rm HT}(x,Q)=F_{2,\rm L}^{\rm TMC}(x,Q)
+H_{2,\rm L}(x)\frac{1~{\rm GeV}^2}{Q^2},
\end{displaymath}  
where $F_{2,\rm L}^{\rm TMC}$ are the LT contributions 
obtained as a result of 
integration of Eqn.~(\ref{eqn:dglap}) with the account of target 
mass corrections \cite {TMC}; $s$ is the total c.m.s. energy; 
$m_{\rm l}$ is the scattered lepton mass; and 
$y$ is the regular lepton scattering variable.
The values of functions $H_{2,\rm L}(x)$ at 
$x=0.3,0.4,0,5,0.6,0.7,0.8$ were fitted,
between these points $H_{2,\rm L}(x)$ were linearly interpolated.
The functions $H_2$ for proton and deuterium
were fitted independently, while the functions $H_L$
for proton and deuterium due to limited accuracy of data, turned out to be 
compatible within errors and all fits were performed 
under constraint $H_{\rm L}^{\rm p}(x)=H_{\rm L}^{\rm d}(x)$.

\begin{figure}[t]
\centerline{\psfig{figure=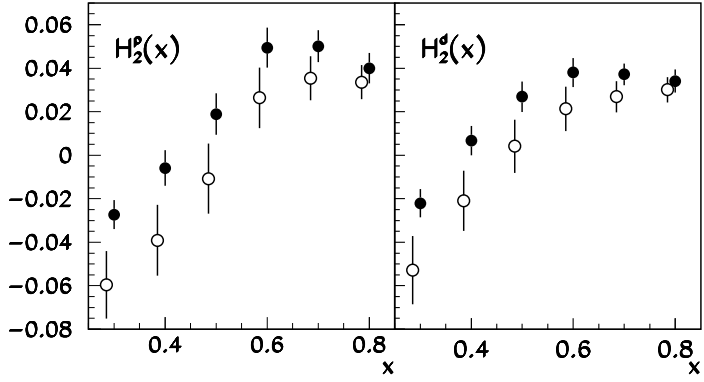,height=5cm}}
\caption{The values of 
proton and deuterium $H_2(x)$ for different values of $k_{\rm R}$
(open circles: $k_{\rm R}=1/2$; full circles: $k_{\rm R}=1$.}
\label{fig:renht}
\end{figure}

\begin{figure}[t]
\centerline{\psfig{figure=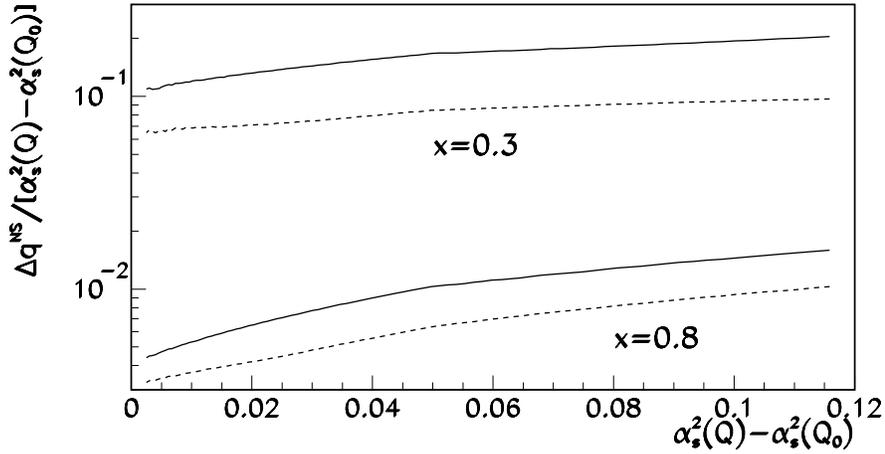,height=6cm}}
\caption{The dependence of proton $\Delta q^{\rm NS}$ on $\alpha_{\rm s}^2$
at $Q_0^2=9$~GeV$^2$ (full lines: $\log_2k_{\rm R}=-1$; 
dashed lines: $\log_2k_{\rm R}=-1/2$).}
\label{fig:f2_nlo}
\end{figure}

\begin{figure}[t]
\centerline{\psfig{figure=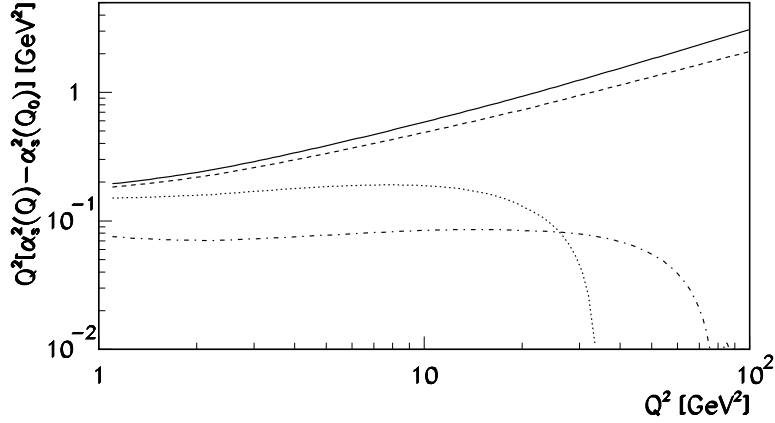,height=6cm}}
\caption{The dependence of 
$Q^2\left[\alpha_{\rm s}^2(Q)-\alpha_{\rm s}^2(Q_0)\right]$
on $Q^2$ (full line: $\alpha_{\rm s}\left(Q_0\right)=0$;
dashed line: $\alpha_{\rm s}\left(Q_0\right)=0.1$;
dotted line: $\alpha_{\rm s}\left(Q_0\right)=0.2$). The dashed-dotted line
corresponds to $Q^2\left[\alpha_{\rm s}^3(Q)-\alpha_{\rm s}^3(Q_0)\right]$
for $\alpha_{\rm s}\left(Q_0\right)=0.18$.}
\label{fig:qalp}
\end{figure}

\begin{figure}[t]
\centerline{\psfig{figure=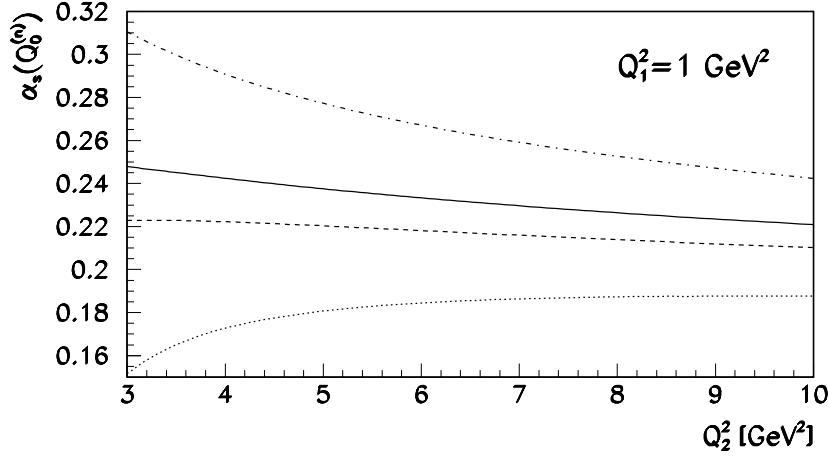,height=6cm}}
\caption{The value of $\alpha_{\rm s}$ at
starting evolution point $Q_0^{(n)}$, that provides 
the power-behaviour simulation of factors
$\left[\alpha_{\rm s}^n(Q^2)-\alpha_{\rm s}^n(Q_0)\right]$
in the region from $Q_1$ to $Q_2$ (full line: $n=1$,
dashed line: $n=2$, dotted line: $n=3$). The value of 
$\alpha_{\rm s}\left(Q_2\right)$ is also given for comparison (dashed-dotted line).}
\label{fig:q0opt}
\end{figure}

\begin{figure}[t]
\centerline{\psfig{figure=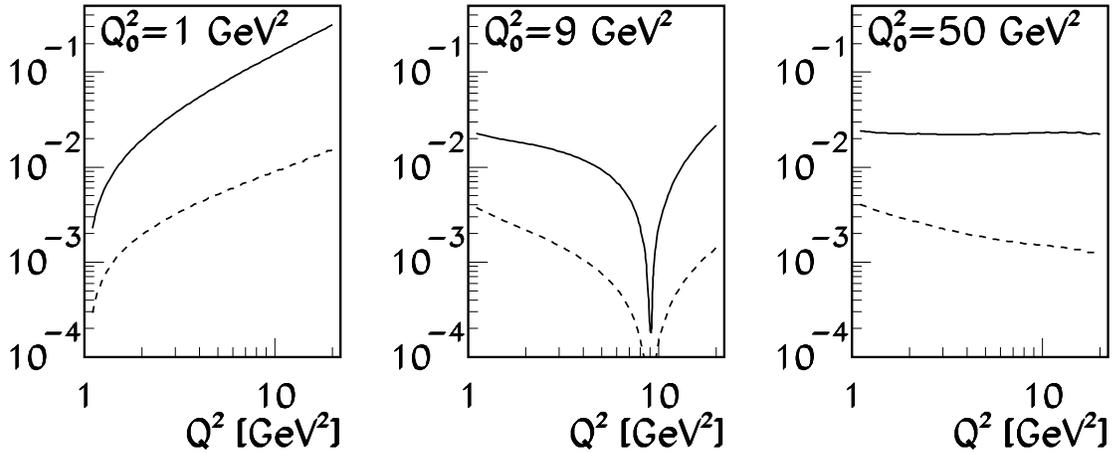,height=6cm}}
\caption{The dependence of 
$\mid Q^2\Delta F_2^{\rm p}\mid [{\rm GeV}^2]$ 
on $Q^2$ at $k_{\rm R}=1/2$ and at 
different values of $x$ and $Q_0$ (full lines: $x=0.3$; 
dashed lines: $x=0.8$).}
\label{fig:delf2_q}
\end{figure}

{\bf 3.} The proton and deuterium $H_2(x)$ for $k_{\rm R}=1$ and $k_{\rm R}=1/2$ 
are given in Fig.~\ref{fig:renht}.
One can see that they depend on $k_{\rm R}$ at $x\sim 0.3$ 
and practically does not depend at $x\sim 0.8$.
To give explanation of this behaviour recall the basic 
properties of solutions to the DGLAP evolution equations.  
After linearization on $\ln k_{\rm R}$
Eqn.~(\ref{eqn:dglap}) can be analytically solved in the Mellin momentum space:
\begin{displaymath}
M^{\rm NS}(n,Q)=M^{\rm NS}(n,Q_0)M_{\rm NLO}(n,\alpha)
\exp\left[g(n)\left(\alpha^2-\alpha^2_0\right)\ln (k_{\rm R})\right],
\end{displaymath}
where $\alpha\equiv\alpha_{\rm s}(Q)$, $\alpha_0\equiv\alpha_{\rm s}\left(Q_0\right)$; 
$M^{\rm NS}(n,Q)$ are Mellin moments of $q^{\rm NS}$; $M_{\rm NLO}(n,\alpha)$ 
defines NLO evolution of these moments;
$g(n)$ is a linear function of Mellin moments of 
the splitting functions $P_{\rm qq}^{\rm NS,(0)}$ and 
$P_{\rm qq}^{\rm NS,(1)}$. Introduce a function 
$\Delta q^{\rm NS}(k_{\rm R})
\equiv q^{\rm NS}(k_{\rm R})-q^{\rm NS}(k_{\rm R}=1)$
that is convenient to study RS dependence. 
The Mellin moments of this function $M^{\Delta}(n,Q)$ are given by 
\begin{equation}
M^{\Delta}(n,Q)=M^{\rm NS}(n,Q_0)M_{\rm NLO}(n,\alpha)
\left\{\exp\left[
g(n)\left(\alpha^2-\alpha^2_0\right)\ln (k_{\rm R})\right]-1\right\}.
\label{eqn:mom}
\end{equation}
In Fig.~\ref{fig:f2_nlo} the precise dependence of 
proton $\Delta q^{\rm NS}$ on 
$\alpha_{\rm s}^2(Q)$ for different $k_{\rm R}$, obtained as the result of
numerical integration of Eqn.~(\ref{eqn:dglap}) at $Q^2_0=9~$GeV$^2$,
is given\footnote{Here and below we do 
not give the results for deuterium since they are similar to the proton ones.}. 
The range of $\alpha_{\rm s}$ in the figure correspond to 
the variation of $Q^2$ from 1 to 9~GeV$^2$, i.e. the region where
the HT contribution is most significant.
It is evident that for all values of $x$ in question 
$\Delta q^{\rm NS}$ is approximately proportional to
$\ln k_{\rm R}$. This means that the 
linearization of Eqn.~(\ref{eqn:dglap}) is justified and
that in Eqn.~(\ref{eqn:mom}) the part of exponent
containing $\ln k_{\rm R}$ can be expanded, so that 
\begin{equation}
\Delta q^{\rm NS}(x,Q)\approx
\ln (k_{\rm R})\left(\alpha^2-\alpha_0^2\right)
\tilde{q}_{\rm NLO}(x,\alpha),
\label{eqn:exp}
\end{equation}
where $\tilde{q}_{\rm NLO}(x,\alpha)$ is the Mellin inverse of   
the product $M^{\rm NS}(n,Q_0)M_{\rm NLO}(n,\alpha)g(n)$.

The $Q$-behaviour of $\tilde{q}_{\rm NLO}$ is defined by $M_{\rm NLO}(n,\alpha)$.
At $x=0.3$ the function $\tilde{q}_{\rm NLO}$ weakly depends on $\alpha$
(see Fig.~\ref{fig:f2_nlo}). This is consequence of 
the well known effect, that the non-singlet QCD evolution 
has stationary point at $x\approx 0.1$
due to the fermion conservation. In the vicinity of stationary point 
scaling violation is small, but, due to at large $n$ the function 
$M_{\rm NLO}(n,\alpha)$ rises with $\alpha$ faster, than at low ones,
at large $x$ the scaling violation is more pronounced and the function  
$\Delta q^{\rm NS}$ rises with $\alpha$ significantly faster than $\alpha^2$.

\begin{figure}[t]
\centerline{\psfig{figure=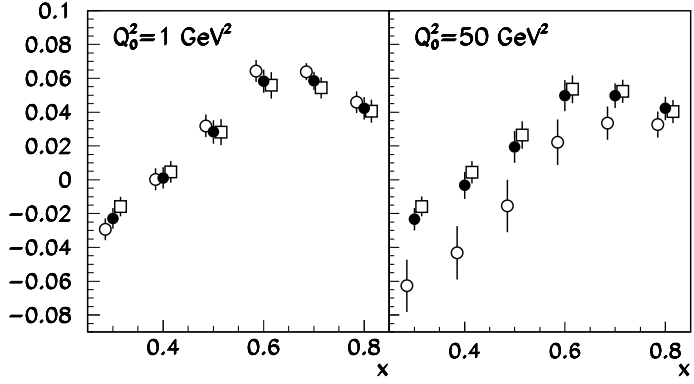,height=5cm}}
\caption{The values of $H^{\rm p}_2(x)$
for different choices of $Q_0$ and RS (open circles: $k_{\rm R}=1/2$;
full circles: $k_{\rm R}=1$; squares: $k_{\rm R}=2$.}
\label{fig:renht_q2}
\end{figure}

In the NLO non-singlet approximation the LT contribution to $F_2$ is given
by 
\begin{displaymath}
F_2^{\rm LT}=q^{\rm NS}+\frac{\alpha_{\rm s}}{2\pi}q^{\rm NS}\otimes C_2^{\rm NS,(1)},
\end{displaymath}
where $C_2^{\rm NS,(1)}$ is the NLO coefficient function.
Since the second term of above expression is suppressed with respect to 
the first one at moderate $x$,
the $Q$-behaviour of function $\Delta F_2(k_{\rm R})\equiv 
F_2^{\rm LT}(k_{\rm R})-F_2^{\rm LT}(k_{\rm R}=1)$ at $x=0.3$
approximately coincides with 
the ones of $\Delta q^{\rm NS}$. The $Q$-behaviour of 
the factor $\left[\alpha_{\rm s}^2(Q^2)-\alpha_{\rm s}^2(Q_0)\right]$, 
coming to Eqn.~(\ref{eqn:exp}) depends on $Q_0$.
If $Q\ll Q_0$ this factor is $\sim 1/\ln^2Q$, i.e.
falls with $Q$ slower, than $1/Q^2$. Meanwhile 
in the vicinity of $Q_0$, where this factor vanishes,
its $Q$-dependence is steeper and it can simulate $1/Q^2$ behaviour in 
a rather wide range of $Q$ (see Fig.~\ref{fig:qalp}). 
One can easily show that the value of $\alpha_{\rm s}$ at a starting evolution 
point $Q_0^{(n)}$, that provides the $1/Q^2$ behaviour simulation of 
the factor $\left[\alpha_{\rm s}^n(Q^2)-\alpha_{\rm s}^n(Q_0)\right]$
in the region from $Q_1$ to $Q_2$, is 
\begin{equation} 
\alpha_{\rm s}\left(Q_0^{(n)}\right)=
\left [\frac{Q_2^2\alpha_{\rm s}^n(Q_2^2)-Q_1^2\alpha_{\rm s}^n(Q_1^2)}
{Q_2^2-Q_1^2}\right ]^{1/n}.
\label{eqn:q0opt}
\end{equation} 
The dependence of $\alpha_{\rm s}\left(Q_0^{(2)}\right)$ on $Q_2$ for 
$Q_1=1~{\rm GeV}^2$
obtained with the help of Eqn.~(\ref{eqn:q0opt}) is given 
in Fig.~\ref{fig:q0opt}. 
For the $Q^2$ interval from 1 to $\sim 10$~GeV$^2$, where the HT are mostly 
significant, $\alpha_{\rm s}\left(Q_0^{(2)}\right)\sim0.2$, that corresponds to 
$\left[Q_0^{(n)}\right]^2\sim 40$~GeV$^2$.
Due to the weak dependence of $\tilde{q}_{\rm NLO}$ 
on $Q$ at small $x$, the function $\Delta F_2$ 
also simulates $1/Q^2$ behaviour at $x=0.3$,
if the starting evolution point is chosen equal to $Q_0^{(2)}$.
At fixed $Q_0$ this simulation, in general, 
becomes less probable with the rise of $x$
just due to the steeper rise of $\Delta q^{\rm NS}$ with 
$\alpha$ at large $x$. 

\begin{figure}[t]
\centerline{\psfig{figure=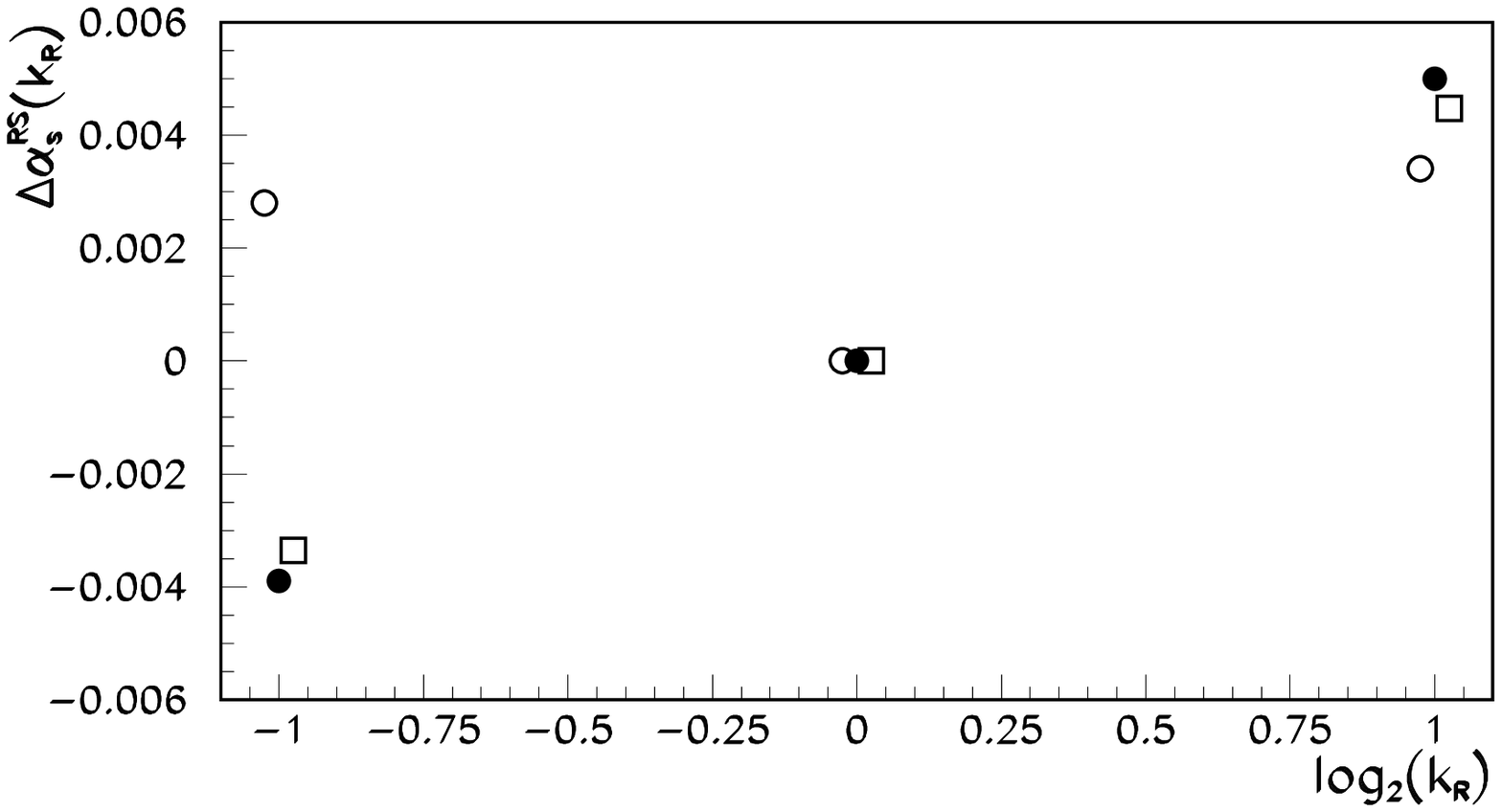,height=6cm}}
\caption{The dependence of 
$\Delta\alpha_{\rm s}^{\rm RS}\left(k_{\rm R}\right)\equiv
\alpha_{\rm s}\left(M_{\rm Z}\right)\mid_{k_{\mathrm R}}-
\alpha_{\rm s}\left(M_{\rm Z}\right)\mid_{k_{\mathrm R}=1}$
on the choice of 
renormalization scale for different $Q_0$ (open circles: $Q_0^2=50$~GeV$^2$, 
full circles: $Q_0^2=1$~GeV$^2$). For comparison are also given
the $\Delta\alpha_{\rm s}^{\rm RS}\left(k_{\rm R}\right)$ values
obtained in the fits  at $Q_0^2=50$~GeV$^2$
with HT fixed at the values obtained in fits with $k_{\rm R}=1$
(squares). Error bars are not given.}
\label{fig:rena_q2}
\end{figure}

The behaviour of $Q^2\Delta F_2^{\rm p}$ at $k_{\rm R}=1/2$ 
for different $x$ and $Q_0$ is given in Fig.~\ref{fig:delf2_q}.
At $Q_0^2=50$~GeV$^2$ and $x=0.3$ the function   
$\Delta F_2$ simulates the $1/Q^2$ behaviour about
perfectly in the total $Q$ region relevant for HT determination.
This leads to the $H_2$ retuning in the fits with different $k_{\rm R}$,
since $\Delta F_2$ is compensated 
by the additional contribution to $H_2$ with the sign opposite
to $\Delta F_2$ (see Fig.~\ref{fig:renht_q2}). 
At $x=0.8$, due to the fall of $\tilde{q}_{\rm NLO}$ with $Q$,
the simulation is much worse. 
If $Q_0^2=1$~GeV$^2$, the absolute value of $Q^2\Delta F_2$ 
steeply rises at $x=0.3$ and small $Q$ due to the factor 
$\left[\alpha_{\rm s}^2(Q^2)-\alpha_{\rm s}^2(Q_0)\right]$.
At $x=0.8$ this rise is not so steep because of 
the fall of $\tilde{q}_{\rm NLO}$ with $Q$,
but it cannot suppress the general rise. As a consequence, 
at small $Q_0$ the function 
$\Delta F_2$ cannot simulate $1/Q^2$ behaviour at all $x$ and
it should be 
at least partially compensated by the change of $Q$-dependence 
of the LT contribution. 
The remnant HT retuning is still possible if the $Q$-dependence of 
$\Delta F_2$ at some $x$ is more similar to $1/Q^2$, than  
to the $Q$-dependence of 
$\partial F_2/\partial \alpha_{\rm s}(M_{\rm Z})$\footnote
{The $Q$-dependence of LT contribution 
is mainly driven by $\alpha_{\rm s}$ and thus the balance between 
the absorbtion of $\Delta F_2$ into the LT contribution and $H_2$
is defined by $\partial F_2/\partial \alpha_{\rm s}(M_{\rm Z})$.}.
The explicit tracing of the  
balance between the $H_2$ retuning and the change of 
LT contribution  is not so simple due to $\alpha_{\rm s}(M_{\rm Z})$
is determined by data at all $x$ and $Q$.
Anyway, as one can see from Fig.~\ref{fig:renht_q2}
at small $Q_0$ the function
$H_2$ is retuned with the change of $k_{\rm R}$
significantly smaller than at large $Q_0$.
At $Q_0^2=9$~GeV$^2$ the retuning effect is almost the same as
for $Q_0^2=50$~GeV$^2$. This is natural since 
the $\Delta F_2$ behaviour at small $Q$ weakly depends on $Q_0$, if  
$Q_0$ is large, and the data are not very sensitive to 
the HT contribution at $Q^2\gtrsim 10$~GeV$^2$. 
Thus the choice of $Q_0$ have small effect on the fitted HT
contribution if $Q^2_0\gtrsim 10$~GeV$^2$.

The RS stability of $\alpha_{\rm s}$ also depends on 
the choice of $Q_0$. In the fit with HT fixed $\Delta F_2$
is compensated by the change of LT contribution, that leads to the shift 
of $\alpha_{\rm s}$. If $H_2$ is released in the fit,
the shift of $\alpha_{\rm s}$ can change due to the  
partial compensation of $\Delta F_2$ by the change of $H_2$.
 Since the $1/Q^2$ simulation of $\Delta F_2$ depends on $Q_0$,  
the $\alpha_{\rm s}$ dependence on $k_{\rm R}$
changes with $Q_0$ as well as $H_2$ (see Fig.~\ref{fig:rena_q2}). 
At large $Q_0$ the $\Delta F_2$ slope on $k_{\rm R}$ is negative
at small $Q$. As a result, in the fit with HT released the $H_2$ slope 
 on $k_{\rm R}$ is positive. Correspondingly the $Q$-dependence 
of LT contribution after HT releasing becomes
weaker at large $k_{\rm R}$ and steeper at small ones, i.e. the 
slope of fitted $\alpha_{\rm s}(M_{\rm Z})$ value on $k_{\rm R}$ decreases
as compared to the fit with HT fixed. In the fit with HT fixed the 
$\alpha_{\rm s}(M_{\rm Z})$ slope on $k_{\rm R}$ is positive and thus
the RS uncertainty on $\alpha_{\rm s}$ at large $Q_0$ becomes smaller.
At small $Q_0$ the $H_2$ slope on $k_{\rm R}$ 
at $x\approx0.5\div0.7$ is negative.
The data from this region of $x$ have the largest impact on
the $\alpha_{\rm s}$ determination 
and thus the $H_2$ releasing leads to the increase
of RS error on $\alpha_{\rm s}$, although the scale of effect is smaller 
as compared with the fit at large $Q_0$
due to the weakness of HT retuning at small $Q_0$. In our analysis
\begin{displaymath}
\alpha_{\rm s}\left(M_{\rm Z}\right)=0.1151\pm0.0015({\rm stat+syst})
\pm0.0045({\rm RS})
\end{displaymath}
for $Q_0^2=1$~GeV$^2$ and 
\begin{displaymath}
\alpha_{\rm s}\left(M_{\rm Z}\right)=0.1183\pm0.0021({\rm stat+syst})\pm0.0013({\rm RS})
\end{displaymath}
for $Q_0^2=9$~GeV$^2$,
where the RS error is estimated as the half of $\alpha_{\rm s}\left(M_{\rm Z}\right)$
spread with the change of $k_{\rm R}$ from 1/2 to 2;
the central value is shifted to the centre of this spread.
The values of $\chi^2$ are approximately the 
same for different $k_{\rm R}$, but
in view of that at small $Q_0$ the total $\alpha_{\rm s}(M_{\rm Z})$
error is about two times larger, than at large one, 
we consider  the $\alpha_{\rm s}$ value 
determined from the fit with large $Q_0$ as more reliable. 

The function $\Delta q^{\rm NS}$, by definition, is connected with the 
NNLO QCD corrections to evolved distributions. A natural assumption is that the 
exponent in NNLO part of moment expression
can be expanded, similarly to the exponent in Eqn.~(\ref{eqn:mom}),
and the NNLO contribution is 
$\sim\left[\alpha_{\rm s}^2(Q^2)-\alpha_{\rm s}^2(Q_0)\right]$
as well as $\Delta q^{\rm NS}$.
One can see from Fig.~\ref{fig:qalp} that the factor
$\left[\alpha_{\rm s}^3(Q^2)-\alpha_{\rm s}^3(Q_0)\right]$ coming to 
N$^3$LO contribution also can simulate $1/Q^2$ behaviour. Moreover, 
the region of $Q$ where the simulation is possible 
widens from the lower orders to the higher ones. 
This gives an indication that the retuning of HT contribution 
after the accounting of HO QCD corrections to DGLAP kernel
would exhibit the same properties as the RS retuning in NLO. 
Note that if $Q_2\gg Q_1$, then $\alpha_{\rm s}\left(Q_0^{(n)}\right)
\sim\alpha_{\rm s}\left(Q_2\right)$ for all $n$
(see Eqn.~\ref{eqn:q0opt} and Fig.~\ref{fig:q0opt}).
This means, that in the analysis with simultaneous determination 
of $\alpha_{\rm s}$ and HT at large $Q_0$ the contribution to $F_2$ due to 
the HO QCD corrections to DGLAP kernel 
can be merely fitted together with the HT, if $g(n)$ and the Mellin moments
of HO splitting functions have the similar large $n$ asymptotes.
The fitting of HO corrections sure leads to the increase of 
$\alpha_{\rm s}$ error\footnote{In the fit
at $Q_0^2=50$~GeV$^2$ with HT fixed the 
$\alpha_{\rm s}\left(M_{\rm Z}\right)$ error is 0.0004
as compared with 0.0021 in the fit with HT released.}.
Meanwhile the reduction of RS uncertainty, that is 
one of the dominant sources of $\alpha_{\rm s}$ error, is larger
and, as one can see from the above,
the total $\alpha_{\rm s}$ error becomes smaller. 

One can see from Fig.~\ref{fig:rena_q2}, that at $Q_0^2=50$~GeV$^2$
the fitted $\alpha_{\rm s}\left (M_{\rm Z}\right)$ value is a 
nonlinear function of $\ln k_{\rm R}$ contrary to   
the fits at $Q_0^2=1$~GeV$^2$. The reason of this difference is 
the large correlation between $\alpha_{\rm s}$ and $H_2$ 
(c.f. Ref.~\cite{ALS}), that depends on how 
well $\partial F_2/\partial \alpha_{\rm s}(M_{\rm Z})$
can simulate $1/Q^2$ behaviour. With the rise of $Q_0$ this 
correlation increases.
For example the correlation coefficient $\rho_{0.5}$
for $\alpha_{\rm s}(M_{\rm Z})$ and $H_2(x=0.5)$ at  
$k_{\rm R}=1/2$ is --0.82 for $Q_0^2=1$~GeV$^2$
and $-0.97$ for $Q_0^2=50$~GeV$^2$. As a consequence, 
at large $Q_0$ a small nonlinearity 
of $q^{\rm NS}$ on $\ln k_{\rm R}$ better manifests 
and takes non-negligible effect on the fitted parameters values
(remind that the effective amplification of nonlinear effects in a fit 
is proportional to $1/(1-\rho^2)$).
For the comparison, the fits with HT fixed exhibit almost linear dependence of 
$\alpha_{\rm s}\left(M_{\rm Z}\right)$ on $\ln k_{\rm R}$
(see Fig.~\ref{fig:rena_q2}).
One of the reflections of this nonlinearity is that
for $Q_0^2=50$~GeV$^2$ the difference between $H_2(x)$ 
at $k_{\rm R}=2$ and at $k_{\rm R}=1$ is small (see Fig.~\ref{fig:renht_q2}).
This is unpleasant feature of the analysis since the variation range
of $k_{\rm R}$ is conventional and the nonlinearity
does not allow to rescale the RS uncertainty. One of the possible ways
to suppress the nonlinear effects is to decrease the correlation between 
HT contribution and $\alpha_{\rm s}$, e.g. adding more data to the analysis. 
From another side this correlation 
leads to the reducing of RS error on $\alpha_{\rm s}$
and it is necessary to keep a balance between the linearity and  
the size of RS error.

\begin{figure}[t]
\centerline{\psfig{figure=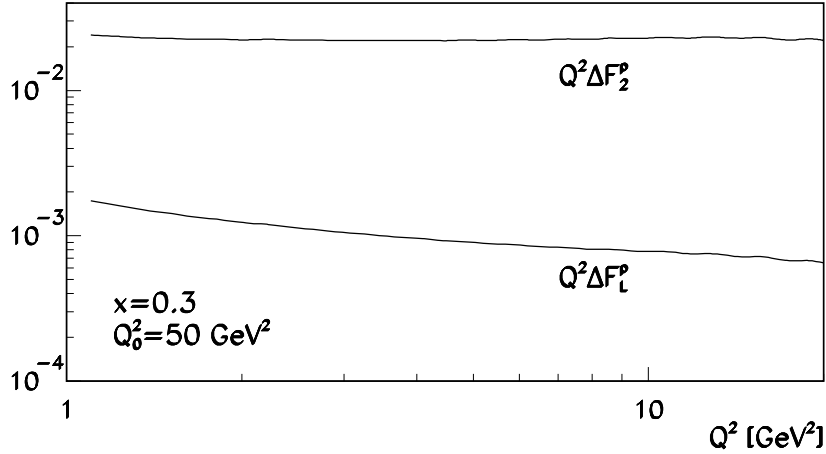,height=6cm}}
\caption{The comparison of $Q^2\Delta F_2^{\rm p}(k_{\rm R}=1/2)$ and 
$Q^2\Delta F_{\rm L}^{\rm p}(k_{\rm R}=1/2)$ dependence on $Q^2$.}
\label{fig:delfl}
\end{figure}

\begin{figure}[t]
\centerline{\psfig{figure=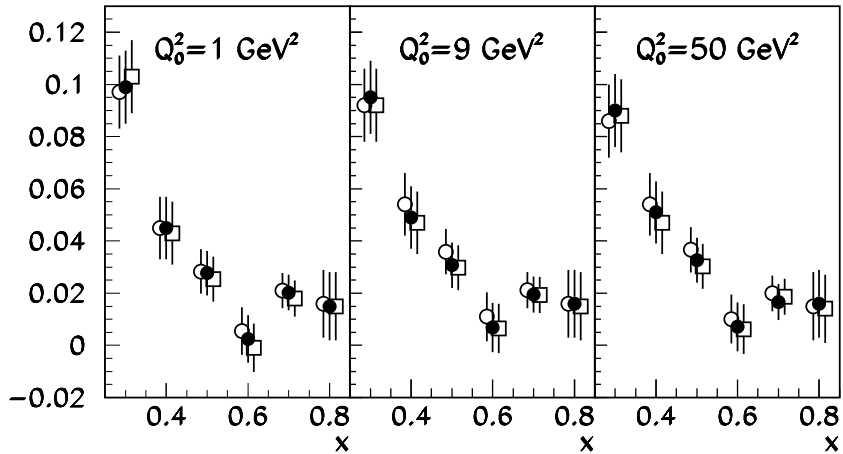,height=6cm}}
\caption{The values of $H_{\rm L}^{\rm p}(x)$
for different choices of $Q_0$ and RS (open circles: $k_{\rm R}=1/2$;
 full circles: $k_{\rm R}=1$; squares: $k_{\rm R}=2$).}
\label{fig:renfl_q}
\end{figure}

 In NLO QCD the LT contribution to structure function $F_{\rm L}$
is proportional to the Mellin convolution 
of $q^{\rm NS}$ with the NLO coefficient function $C_{\rm L}^{\rm NS,(1)}$:
\begin{displaymath}
F_{\rm L}^{\rm LT}=\frac{\alpha_{\rm s}}{2\pi}q^{\rm NS}
\otimes C_{\rm L}^{\rm NS,(1)}.
\end{displaymath}
Due to the convolution smearing a function $\Delta F_{\rm L}(k_{\rm R})\equiv
 F_{\rm L}^{\rm LT}(k_{\rm R})-F_{\rm L}^{\rm LT}(k_{\rm R}=1)$
depends on $Q$ steeper, than $\Delta F_2$.
One can see, that at $x=0.3$ and large $Q_0$ the function
$Q^2\Delta F_{\rm L}(k_{\rm R}=1/2)$ falls in about two times 
in the region of $Q^2=1\div~3~$GeV$^2$ (see Fig.~\ref{fig:delfl}).
At large $x$ and low $Q_0$ it falls even more steeply. 
As a result, $H_{\rm L}(x)$ exhibits only 
weak dependence on $k_{\rm R}$ for all $x$ and $Q_0$
(see Fig.~\ref{fig:renfl_q}).

{\bf 4.} In summary, we can conclude that the HT contribution to  
structure function $F_2$ extracted in the NLO QCD analysis of 
nonsinglet SLAC-BCDMS-NMC data is retuned with the RS change. 
This retuning depends on the choice of starting evolution point
$Q_0$ and $x$. At $Q_0\gtrsim 10$~GeV$^2$ the HT contribution to $F_2$
is retuned at small $x$ and not almost retuned at 
large $x$; at small $Q_0$ it exhibits approximate RS stability for 
all $x$ in question. The RS sensitivity of $\alpha_{\rm s}$ also depends on the 
choice of $Q_0$: at large $Q_0$ this sensitivity is weaker, than at small ones.
The HT contribution to $F_{\rm L}$ is RS stable for all $Q_0$ and $x$.

The RS stability of HT contribution is important for 
clarification of their nature: due to the both HT and 
HO corrections fall with $Q$, it was often 
claimed that the extracted HT terms can contain contribution
from HO. The HT absorbtion 
by NNLO correction was observed in the analysis of 
neutrino structure function $xF_3$ \cite{KKPS}, although the effect
was smashed due to low accuracy of the data.
Our results indicate, that in the analysis of high statistical 
charged leptons DIS data the unambiguous separation of twist-4 contribution
and NNLO QCD corrections to DGLAP kernel is possible if $Q_0$ is low.
This conclusion is especially important because
no complete NNLO calculation of the splitting functions
is available up to now and it is impossible to perform 
exact direct clarification of this point. 

{\bf Acknowledgments}

I am indebted to A.Kataev, L.Mankievitz, and G.Martinelli for valuable 
discussions and comments.

\vskip 2cm
\begin{flushright}
\it Received November ??, 1999
\end{flushright}


\begin{thebibliography}{50}

\bibitem{REW} M. Beneke, CERN-TH/98-233, hep-ph/9807443.

\bibitem{ALS} S.I. Alekhin, Phys. Rev. {\bf D59}, (1999) 114016.

\bibitem{FL}
S.I. Alekhin, IHEP 99-03, hep-ph/9902241; Eur. Phys. J. in print.

\bibitem{MOR99}
S.I. Alekhin, hep-ph/9907350, in the Proceedings of XXXIV 
Moriond conference ``QCD and high energy hadronic interactions'', 
March 20-27, 1999, Les Arcs.

\bibitem {MRS} A.D. Martin, R.G. Roberts, W.J. Stirling, 
Phys. Lett. {\bf B266}, (1991) 273.

\bibitem{VM}
   M. Virchaux, A. Milsztajn, Phys. Lett. {\bf B274}, (1992) 221.

\bibitem{DATA} L.W. Whitlow et al., Phys. Lett. {\bf B282}, (1992) 475;\\
BCDMS collaboration, A.C. Benvenuti et al., Phys. Lett. {\bf B223}, (1989) 485;\\
BCDMS collaboration, A.C. Benvenuti et al., Phys. Lett. {\bf B237}, (1990) 592;\\
NM collaboration, M. Arneodo et al., Nucl. Phys. {\bf B483}, (1997) 3.

\bibitem{AL96} S.I. Alekhin, Eur. Phys. Jour. {\bf C10}, (1999) 395.

\bibitem{TMC} H. Georgi, H.D. Politzer, Phys. Rev. {\bf D14}, (1976) 1829.

\bibitem{KKPS} A.L. Kataev, A.V. Kotikov, G. Parente, A.V. Sidorov,
Phys. Lett. {\bf B417}, (1998) 374.

\end{thebibliography}
\end{document}